\documentclass[hyper, a4paper, 12pt, oneside]{JHEP3}
\usepackage{amssymb,amsmath}
\usepackage{latexsym}

\usepackage{graphicx}

\def\R{\mathbb{R}}

\def\C{\mathbb{C}}

\preprint{APC/Paris7}

\title{Finite dimensional  quantizations of the $(q,p)$ plane :  new space and momentum inequalities.}

 \author{Jean-Pierre Gazeau and Fran\c{c}ois-Xavier Josse-Michaux \\ Laboratoire
Astroparticules et Cosmologie\thanks{UMR 7164 (CNRS,Universit\'e Paris 7, CEA, Observatoire de Paris)}
\\  Boite 7020, Universit\'e Paris 7
Denis DiderotÊ\\ F-75251 Paris Cedex 05, France 
\\ E-mail: \email{gazeau@ccr.jussieu.fr, fixjm@netcourrier.com }} 

 \author{Pascal Monceau\\ Laboratoire Mati\`ere et Syst\`emes Complexes, CNRS UMR 7057\\
 Boite 7020, Universit\'e Paris 7
Denis Diderot \\ F-75251 Paris Cedex 05, France\\ E-mail: \email{pmo@ccr.jussieu.fr}}

\date{\today}

 \abstract{
We present  a  $N$-dimensional quantization \emph{\`a la} Berezin-Klauder or \emph{frame} quantization of the complex plane based on
overcomplete families of states (coherent states) generated by the  $N$ first  
harmonic oscillator eigenstates. The spectra of position and momentum operators are finite and eigenvalues are equal, up to a factor, to the zeros of Hermite polynomials. From numerical and theoretical studies of the  large $N$ behavior of the product $\lambda_m(N)\, \lambda_M(N)$   of  non null smallest positive  and largest eigenvalues, we infer the inequality $  \delta_N(Q)\, \Delta_N(Q) = \sigma_N \overset{<}{\underset{N \to \infty}{\to}} 2 \pi$ (resp.  $ \delta_N(P) \, \Delta_N(P) = \sigma_N \overset{<}{\underset{N \to \infty}{\to}} 2 \pi $) involving, in suitable units,  the minimal ($ \delta_N(Q)$) and maximal ($ \Delta_N(Q)$)  sizes of  regions of space (resp. momentum) which are accessible to exploration within this finite-dimensional quantum framework. Interesting issues on the measurement process and connections with the finite Chern-Simons matrix model for the Quantum Hall effect are discussed.}


\begin{document}


\section{\label{sec:level0} Introduction}

The idea of exploring various aspects of Quantum Mechanics  by restricting the Hilbertian framework to finite-dimensional space   has been increasingly  used in the last decade, mainly in the context of Quantum Optics \cite{buwiknla,kuwazh}, but also in the perpective of non-commutative geometry and ``fuzzy'' geometric objects \cite{kezo}. For Quantum Optics, a comprehensive  review (mainly devoted to the Wigner function) is  provided by Ref.\cite{milei}. In \cite{kuwazh}, the authors defined normalized finite-dimensional coherent states by  truncating the Fock expansion of the standard coherent states.  Besides, basic features of the quantum Hall effect can be described within  the finite matrix Chern-Simons approach \cite{poly}.

It is well known, essentially  since Klauder and Berezin, that one can easily achieve canonical quantization of the classical phase space by using standard coherent states \cite{klau1,ber,KS85,csfks,klau2}. In this paper we apply a related quantization method to the case in which the space of quantum states is finite-dimensional.  Interesting new inequalities concerning observables emerge from this finite-dimensional quantization, in particular in the context of the quantum Hall effect. 

 This coherent state quantization  with its various generalizations reveals itself
as an efficient tool for quantizing physical systems for which the implementation
of more traditional methods is unmanageable (see for instance \cite{Gahulare,alenga,gapie}).
In order to become familiar with our approach, we start  the body of the paper by  presenting in Section \ref{sec:level1} the general mathematical framework, and we apply in Section \ref{sec:level2} this formalism to the elementary example of the motion of the particle on the real line.
We next
consider in Section \ref{sec:level3} finite-dimensional   quantizations.
After working out the algebras of these quantum systems, we shall explore their respective physical
meaning in terms of lower symbols, localisation and momentum range properties. New inequalities are derived in Section \ref{sec:level4}. More precisely, from the existence of a finite spectrum of the position and momentum operators in finite-dimensional quantization, we find that there exists an interesting correlation between the  size $\delta_N$ of the minimal ``forbidden'' cell and the width $\Delta_N$ of the spectrum (``size of the  universe" accessible to measurements from the point of view of the specific system being quantized). This correlation reads in appropriate units $\delta_N \times \Delta_N = \sigma_N$, and numerical explorations, validated by theoretical arguments, indicate that the strictly increasing sequence converges: $\sigma_N \xrightarrow[N \to \infty]{} \sigma \sim 2 \pi$. A similar result holds for the spectra of the momentum operators. In Section \ref{sec:level5}, we  sketch a discussion about the consequences  of our inequalities in term of physical interpretation, particularly in connection with the quantum Hall matrix model.


\section{\label{sec:level1} General setting: quantum processing of a measure space}
In this 
section, we present the method of quantization we will apply in the sequel to a simple model, for instance the motion  of a particle on the line, or more generally a system with one degree of freedom. The method, which is based on coherent states \cite{csfks,csbook} or \emph{frames} \cite{aag1} in Hilbert spaces is inspired by previous approaches proposed by Klauder \cite{klau1,klau2}  and Berezin \cite{ber}. More details and examples concerning the method can be found  in the references  \cite{Gahulare,alenga,gapie}.

Let us start with an arbitrary measure space  $(X, \mu)$. This set might be a classical
phase space, but actually it can be any
set of data accessible to observation. The existence of a measure provides
us with a statistical reading of the set of measurable real- or
complex-valued functions $f(x)$ on $X$: computing for instance
average values on subsets with bounded measure. Actually, both
approaches deal with quadratic mean values and correlation/convolution
involving  pairs of functions, and the natural framework of studies is the
complex (Hilbert) spaces, $L^2(X, \mu)$
 of square integrable
functions $f(x)$ on $X$: $\int_X \vert
f(x)\vert^2 \, \mu(dx) < \infty $. One will speak of {\it
finite-energy} signal in Signal Analysis and of (pure) quantum state in
Quantum Mechanics. However, it is precisely at this stage that
``quantum processing'' of $X$ differs from signal processing on at
least three points: 
\begin{enumerate}
\item not all square integrable functions are eligible as quantum
states, 
\item a quantum state is defined up to a nonzero factor,
\item those ones among functions $f(x)$ that are eligible as quantum
states with  unit norm, $\int_X \vert f(x)\vert^2\, \mu(dx)
= 1$, give rise to a probability interpretation : $X \supset \Delta
\rightarrow \int_{\Delta} \vert f(x) \vert^2 \mu(dx)$ is a
probability measure interpretable in terms of localisation in the
measurable $\Delta$. This is  inherent to the computing of mean values of
quantum observables, (essentially) self-adjoint operators with domain
included in the set of quantum states. 
\end{enumerate}
The first point lies at the heart of the {\it quantization} problem:
what is the more or less canonical procedure allowing to select
quantum states among simple signals? In other words, how to select
the right (projective) Hilbert space  ${\mathcal H}$, a closed
subspace of $L^2(X, \mu)$, (resp.  some isomorphic copy of it) or equivalently the corresponding
orthogonal projecteur $I_{{\mathcal H}}$ (resp. the identity operator)?

In various circumstances, this question is answered through the
selection, among elements of $L^2(X, \mu)$, of an orthonormal set
$\mathcal{S}_N = \{ \phi_n(x) \}_{n = 0}^{N-1}$, $N$ being finite or
infinite, which spans, by definition,  the separable Hilbert subspace
${\mathcal H}
\equiv {\mathcal H}_N$. The crucial point is that these elements have to fulfill
the following condition :
\begin{equation}{\cal N} (x) \equiv \sum_n \vert \phi_n (x) \vert^2 <
\infty \ \mbox{almost everywhere}. \label{factor}
\end{equation}
Of course, if $N \geq 1$ is finite the above condition is trivially checked.

We now consider the   family of states $\{ | x \rangle \}_{x\in X}$ \underline{in} ${\mathcal H}_N$  obtained through the
following linear superpositions: \begin{equation}
| x\rangle \equiv \frac{1}{\sqrt{{\mathcal N} (x)}} \sum_n \overline{\phi_n} (x)
| \phi_n\rangle, \label{cs}
\end{equation}
in which the ket $| \phi_n\rangle$ designates the element $\phi_n(x)$ in a
``Fock'' notation and $\overline{\phi_n} (x)$ is the complex conjugate of $\phi_n(x)$. 
This defines an injective map
\begin{equation}
  X \ni x
\rightarrow | x \rangle \in {\mathcal H}_N,    
\end{equation}
and the above Hilbertian superposition makes sense provided that set $X$ is equipped of a mild topological structure for which this map is continuous. It is not difficult to check that states
(\ref{cs}) are \textit{coherent} in the sense that they obey
the following two conditions:
\begin{itemize}
\item {\bf Normalisation}
\begin{equation}\langle \, x\, | x \rangle = 1, \label{norma}
\end{equation}\item {\bf Resolution of the unity in ${\mathcal H}_N$}
\begin{equation}\int_X | x\rangle \langle x  | \,\, \nu(dx)=
I_{{\mathcal H}_N}, \label{iden}
\end{equation}
where $\nu(dx) = {\mathcal N}(x)\,\mu(dx)$ is another
measure on $X$, absolutely continuous with respect to $\mu(dx)$. The coherent
states (\ref{cs}) form in general an overcomplete (continuous) basis of ${\mathcal H}_N$.
Actually, the term of \emph{frame} \cite{aag1} is more appropriate for designating the total family $\{ | x\rangle \}_{x \in X} $.
\end{itemize}

The resolution of the unity in ${\mathcal H}_N$ can alternatively
be understood in terms of the scalar product $\langle \, x\, | x'
\rangle$ of two states of the family. Indeed, (\ref{iden}) implies
that, to any vector $| \phi \rangle$ in ${\mathcal H}_N$ one can
isometrically associate the function \begin{equation}
\phi(x) \equiv
\sqrt{\mathcal N(x)}\langle x\, | \phi \rangle
\end{equation} in $L^2(X, \mu)$, and
this function obeys \begin{equation}
\phi(x) = \int_X \sqrt{\mathcal
N(x)\mathcal N(x')} \langle x| x' \rangle \phi(x')\, \mu(dx') .
\end{equation} Hence, ${\mathcal H}_N$ is isometric to a reproducing
Hilbert space with kernel \begin{equation}
{\cal K}(x,x') = \sqrt{\mathcal
N(x)\mathcal N(x')}
\langle x\, | x' \rangle,
\end{equation} and the latter assumes finite diagonal
values ({\it a.e.}), ${\cal K}(x,x) = \mathcal N(x)$, by construction.

A {\it classical} observable is  a
function $f(x)$ on $X$ having specific properties in relationship
with some supplementary structure allocated to $X$, namely topology,
geometry .... Its  quantization   simply consists in 
associating to $f(x)$
the operator
\begin{equation}A_f := \int_X f(x) | x\rangle \langle x| \, \nu(dx).
\label{oper}
\end{equation}In this context, $f(x)$ is said upper (or
contravariant) symbol of the operator $A_f$ and denoted by $f = \hat{A}_f $, whereas the mean value
$\langle x| A | x\rangle$ is said lower (or covariant) symbol of an operator
$A$ acting on ${\mathcal H}_N$ \cite{ber} and denoted by $\check{A}_f$.
Through this approach, one can say
that a quantization of the observation set is
in one-to-one correspondence with the choice of a frame in the sense
of (\ref{norma}) and (\ref{iden}). To a certain extent, a
quantization scheme consists in adopting a certain point of view in
dealing with $X$. This frame can be discrete, continuous, depending
on the topology furthermore allocated to the set $X$, and it can be
overcomplete, of course. The validity of a precise frame choice with regard to a certain physical context  is
asserted by comparing spectral characteristics of quantum observables
$A_f$ with experimental data.


\section{\label{sec:level2}The standard case}

Let us illustrate the above construction with the well-known Klauder-Glauber-Sudarshan coherent
states \cite{csfks}. The observation set $X$ is the classical phase space $\R^2 \simeq \C = \{ x
\equiv z = 
\dfrac{1}{\sqrt{2m\omega u_A}}(m\omega q+ip) \}$ (in complex notations) of a system with one degree of freedom and experiencing a motion with characteristic time $\omega^{-1}$ and action $u_A$. Note that the characteristic length and momentum of this system are $l_c = \sqrt{\dfrac{u_A}{m\omega}}$ and  $p_c = \sqrt{m\omega u_A}$ respectively, whereas the phase-space variable $z$ can be expressed in   units of square root of action $\sqrt{u_A}$. Now, we could as well deal with an oscillating system like a biatomic molecule. Of course, in the domain of validity of quantum mechanics, it is natural to choose $u_A = \hbar$.  The
measure on $X$ is gaussian, $\mu(dx) = \frac{1}{\pi}\, e^{-\frac{\vert z \vert^2}{u_A}}\, d^2 z $ where $d^2
z$ is the Lebesgue measure of the plane. In the sequel, we shall work in suitable units, \emph{i.e.} with $m = 1$, $\omega = 1$, and $u_A = 1$. 

The functions $\phi_n (x)$ are the normalised
powers of the conjugate of the complex variable $z$, $\phi_n (x) \equiv \frac{\bar{z}^n}{\sqrt{n!}}$, so that the
Hilbert subspace ${\cal H}$ is the so-called Fock-Bargmann space of all anti-entire functions that are
square integrable with respect to the gaussian measure. Those states are eigenvectors of  the number
operator
$\mathfrak{N}$ which is identical to the dilation operator $\mathfrak{N}=z \frac{\partial}{\partial z}$. Since $
 \sum_n \frac{\vert z \vert^{2n}}{n!} =  e^{\vert z \vert^2}$, the coherent states read
\begin{equation}| z\rangle = e^{-\frac{\vert z \vert^2}{2}} \sum_n  \frac{z^n}{\sqrt{n!}}| n\rangle,
\label{scs}
\end{equation} where we have adopted the usual notation $| n \rangle = |\phi_n\rangle$. 

One easily checks the normalisation and unity resolution:
\begin{equation}\langle z\, | z \rangle = 1,\  \ 
\frac{1}{\pi}\int_{\C}  | z\rangle \langle z| \, d^2 z= I_{{\cal H}},
\label{pscs}
\end{equation}Note that the reproducing kernel is simply given by $e^{\bar{z}z'}$.
The quantization of the observation set is hence achieved by selecting in the original Hilbert
space $L^2(\C, \frac{1}{\pi}e^{-\vert z \vert^2}\, d^2 z)$ all anti-holomorphic entire functions, which geometric
quantization specialists would call a choice of polarization.  Quantum operators acting on ${\cal
H}$ are yielded by using (\ref{oper}). We thus have for the most basic one,
\begin{equation}\frac{1}{\pi}\int_{\C}  z\, | z\rangle \langle z| \, d^2 z = \sum_n \sqrt{n+1} 
| n\rangle \langle n+1| \equiv a,
\label{low}
\end{equation}which is the lowering operator, $a | n\rangle = \sqrt{n} | n - 1\rangle$. Its adjoint
$a^{\dagger}$ is obtained by replacing $z$ by $\bar{z}$ in (\ref{low}), and we get the
factorisation $\mathfrak{N} = a^{\dagger}a$ together with the  commutation rule $\lbrack a, a^{\dagger}
\rbrack = I_{{\cal H}}$. Also note that $ a^{\dagger}$ and $a$ realize on ${\cal H}$ as
multiplication operator and derivation operator respectively, $a^{\dagger}f(z) = zf(z), \ af(z)
= df(z)/dz$. From $q = \frac{1}{\sqrt{2}}(z +
\bar{z})$ and  $p = \frac{1}{i\sqrt{2}}(z - \bar{z})$, one easily infers by linearity that $q$ and $p$
are   upper symbols for $\frac{1}{\sqrt{2}}(a + a^{\dagger}) \equiv Q$ and $\frac{1}{i\sqrt{2}}(a - a^{\dagger})
\equiv P$ respectively. In consequence,  the self-adjoint operators $Q$ and $P$ obey the canonical
commutation rule $\lbrack Q, P \rbrack = iI_{{\cal H}}$, and for this reason fully deserve  the
name of position and momentum operators of the usual (galilean) quantum mechanics, together with
all  localisation properties specific to the latter. 

These standard states have many interesting properties. Let us recall two of them: they are
eigenvectors of the lowering operator, $a | z\rangle = z | z\rangle$, and they
saturate the Heisenberg inequalities : $\Delta Q\, \Delta P = \frac{1}{2}$.  It should be
noticed that they  also  pertain to the group theoretical construction since they are obtained from
unitary Weyl-Heisenberg transport of the ground state: $ | z\rangle = \exp (z
a^{\dagger} -   \bar{z} a)| 0\rangle$.


\section{\label{sec:level3}The finite-dimensional quantization}
Let us now consider the generic orthonormal set with $N$ elements:
\begin{equation}
\label{ortN}
\phi_0(x)=1, \phi_1(x)=\bar{z},\dotsc  \phi_{N-1}(x)=\frac{\bar{z}^{(N-1)}}{\sqrt{(N-1)!}}.
\end{equation}

 The coherent states read :
\begin{equation}
\label{cs3}
|z\rangle= \frac{1}{\sqrt{\mathcal{N}(x)}} \sum_{n = 0}^{N-1} \frac{ z^n}{\sqrt{n!}} |n \rangle,
\end{equation}
with
\begin{equation}
\label{normN}
\mathcal{N}(x)=\sum_{n = 0}^{N-1} \frac{\vert z \vert^{2n}}{n!}.
\end{equation}

They provide the following  quantization of the classical position $q$ and momentum $p$ :
\begin{equation}
\frac{1}{\pi}\int_{\C} \left\lbrace\begin{array}{c}
      q    \\
      p  
\end{array}\right\rbrace  | z\rangle \langle z| \mathcal{N}(x) e^{-\vert z \vert^2}\, d^2 z = 
 \left\lbrace\begin{array}{c}
      Q_N    \\
      P_N 
\end{array}\right\rbrace\label{finquant}
\end{equation}
Matrix elements of the position operator $Q_N$ and momentum  operator $P_N$ are given by
\begin{equation}
\label{posN}
Q_N(k,l)=\frac{1}{\sqrt{2}}\,(\sqrt{k}\,\delta_{k,l-1}\,+\sqrt{k-1}\,\delta_{k,l+1}\,),
\end{equation}
\begin{equation}
\label{momN}
P_N(k,l)=-i\,\frac{1}{\sqrt{2}}\,(\sqrt{k}\,\delta_{k,l-1}\,-\sqrt{k-1}\,\delta_{k,l+1}\,) ,
\end{equation}
for $\ 1\leq k,l \leq N$.
Their commutator is ``almost'' canonical:
\begin{equation}
\label{cacomN}
[Q_N,P_N] = iI_N - i N E_N,
\end{equation}
where $E_N$ is the orthogonal projector on the last basis element,
\begin{equation*}
\label{ }
E_N = \left(\begin{array}{ccc}0 &  \dotso & 0 \\ \vdots & \ddots & \vdots \\0 & \dotso & 1\end{array}\right).
\end{equation*}
The appearing of such a projector in (\ref{cacomN}) is clearly a consequence of the truncation at the $N^{\mathrm{th}}$ level.
We shall study the spectra of these operators in the next section.

The corresponding truncated harmonic oscillator hamiltonian $H_N = \dfrac{1}{2}(P_N^2 + Q_N^2)$ is diagonal with matrix elements :
\begin{equation}
\label{hamN}
H_N(k,l)=\frac{1}{2}\,(2k-1-N\delta_{k,N})\,\delta_{k,l}.
\end{equation}
Since $H_N$  is diagonal, its eigenvalues are trivially  $\frac{1}{2}\,(2k-1-N\delta_{N,k})$ and are identical to the lowest  eigenenergies of the harmonic oscillator, except for the $N^{\mbox{th}}$ one which is equal to
 $\dfrac{N-1}{2}$ instead of  $N-\dfrac{1}{2}$.
One should notice that its nature differs according to the parity of $N$: it is degenerate if $N$ is even
since then $\dfrac{N}{2} - \dfrac{1}{2}$ is already present in the spectrum
 whereas it assumes the intermediate value $\left\lfloor \dfrac{N}{2} \right\rfloor$ between two expected values if $N$ is odd. 
 
 Let us now consider the mean values or lower symbols of the position and momentum operators. We find:
\begin{equation}
\label{mvalQP}
\langle z|Q_N|z \rangle=C(\vert z \vert )q,  \  \langle z|P_N|z \rangle=C( \vert z \vert)p,
\end{equation}
where the corrective factor
\begin{equation}
C(\vert z \vert)=\frac{1}{\mathcal N (z)}
\sum_{j=1}^{N-1}\frac{(\vert z \vert)^{2(j-1)}}{(j-1)!}
\end{equation}
goes to $1$ as $N \to \infty$.

Lower symbols of the operators 
 $Q_N^2$, $P_N^2$ and $H_N$ are given by:
 \begin{equation}
\label{mvalsq}
\langle z|\left\lbrace\begin{array}{l}
 Q_N^2        \\
  P_N^2       
\end{array}\right\rbrace|z\rangle = A(\vert z \vert) \pm B(\vert z \vert),\  \langle z|H_N |z\rangle = A(\vert z \vert),
\end{equation}
where
\begin{align*}
A(\vert z \vert) &=  \frac{1}{\mathcal N(z)}\sum_{k=1}^{N}\frac{\vert
z \vert^{2(k-1)}}{(k-1)!}\,\left(\frac{2k-1-N\delta_{N,k}}{2}\right),\\
B(\vert z \vert) &= \frac{1}{\mathcal N(z)}
\sum_{k=1}^{N-2}\frac{\vert
z\vert^{2(k-1)}}{(k-1)!}\,\frac{z^2+\bar{z}^2}{2}.
\end{align*}

The behavior of these lower symbols in (\ref{mvalsq}) in function of $(q,p)$, with the particular value $N = 12$, is shown in Fig. \ref{Q2P2}. One can see that these mean values are identical, albeit the lower symbol  of $P_N^2$ is obtained from that of $Q_N^2$ through a rotation by $\frac{\pi}{2}$ in the complex plane.


\begin{figure}
\begin{center}
\includegraphics[width=6cm, angle=270]{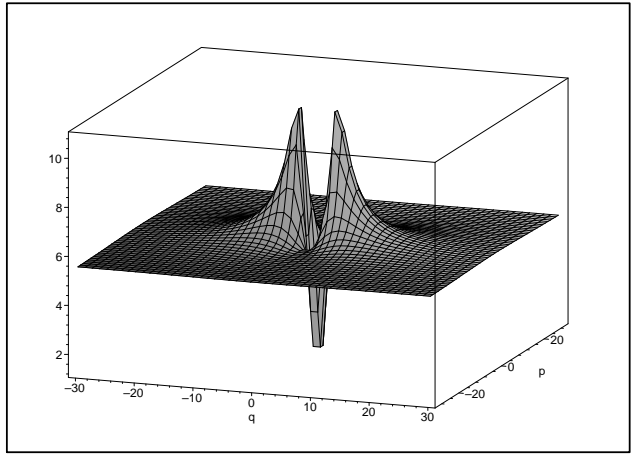}
\caption{$(q,p)$ behavior of the meanvalues (lower symbols) of the operator $Q_N^2$  in the coherent state $|z\rangle$ for $N = 12$.}
\label{Q2P2}
\end{center}
\end{figure}



\begin{figure}
\begin{center}
\includegraphics[width=6cm, angle=270]{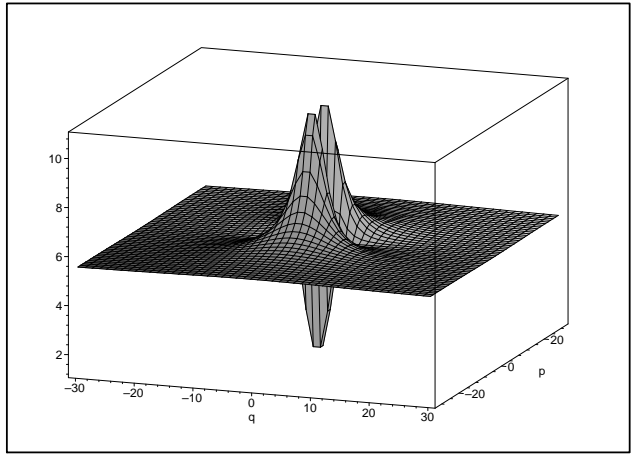}
\caption{$(q,p)$ behavior of the meanvalues (lower symbols) of the operator  $P_N^2$ in the coherent state $|z\rangle$ for $N = 12$.}
\label{Q2P2}
\end{center}
\end{figure}

\begin{figure}
\begin{center}
\includegraphics*[angle=270,width=8cm]{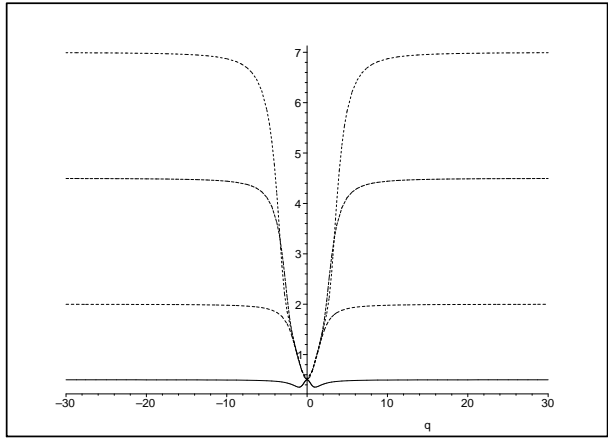}
\caption{Behavior of $\Delta Q_N\,\Delta P_N$ in function of $q$, at $p = 0$, for different values of $N$, $N = 2$ (lowest curve), $5, 10,15$ (upper curve).}
\label{DQDP}
\end{center}
\end{figure}

From all these meanvalues we can deduce the product $\Delta Q_N\,\Delta P_N$, where $\Delta Q_N=\sqrt{\langle z |Q_N^2|z \rangle-(\langle z|Q_N|z
\rangle)^2}$. Due to rotational invariance, it is enough to consider its behavior in function of $q$, at $p = 0$, as  is shown in Fig. \ref{DQDP} for different values of $N$, $N = 2, 5, 10, 15$. One can observe that $\Delta Q_N\,\Delta P_N = 1/2$, \emph{i.e.} the product assumes, at the origin of the phase space the minimal value it would have in the infinite-dimensional case (with $\hbar = 1$). Note that, for the minimal case $N=2$, the value $1/2$ is a supremum (!), and the latter is reached for almost all values of $z$ except in the range  ${\vert z\vert \lesssim 10}$. For higher values of $N$, there exists around the origin a range  of values of $\vert z \vert$, where the product is equal to $\frac{1}{2}$. This range increases with $N$ as expected since the Heisenberg inequalities are saturated with standard coherent states ($N = \infty$). 


\begin{figure}
\begin{center}
\includegraphics*[angle=270,width=8cm]{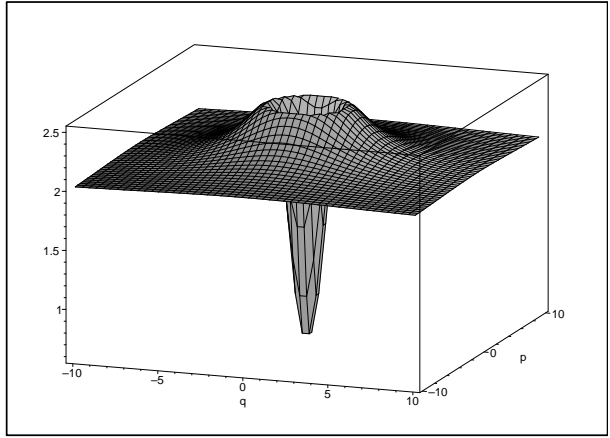}
\caption{ Meanvalue $ \langle z|H_N|z\rangle$ of the  harmonic oscillator hamiltonian as a function of $z = \frac{1}{\sqrt{2}}(q + i p)$ for $N=5$.}
\label{mvH}
\end{center}
\end{figure}

\begin{figure}
\begin{center}
\includegraphics*[angle=270,width=8cm]{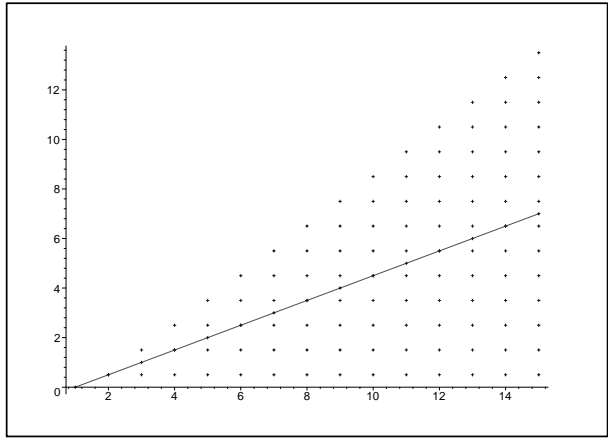}\label{tron}
\caption{Spectrum of the harmonic oscillator Hamiltonian $H_N$ in function of $N$. One clearly  sees the appearance of a degeneracy or an intermediate value instead in the vicinity of the middle of the spectrum,  according to the parity of $N$, along the dotted line.
}
\label{spH}
\end{center}
\end{figure}

Let us finally consider the behavior in function of $\vert z\vert$ of  the lower symbol of the harmonic oscillator Hamiltonian given in Eq. (\ref{mvalsq}).  From Figs. \ref{mvH} and 
\ref{spH} in which are shown respectively the meanvalue of $H_N$  at $N = 5$, and  the energy spectrum for different values of $N$, one can see the influence of truncating the dimension of the space of states.


\section{\label{sec:level4}Localization and momentum of the finite-dimensional quantum system}

We now examine the spectral features of the position and momentum operators $ Q_{N}$, $ P_{N}$ given in the $N$-dimensional case by
Eqs.(\ref{posN}) and (\ref{momN}), \emph{i.e.} in explicit matrix form by:

\begin{equation}
\label{posN}Q_{N} = \left(
\begin{array}
[c]{ccccc}%
0 & \frac{1}{\sqrt{2}} & 0 & \dotso & 0\\
\frac{1}{\sqrt{2}} & 0 & 1 & \dotso & 0\\
0 & 1 & \ddots & \ddots & \vdots\\
\vdots & \dotso & \ddots & 0 & \sqrt{\frac{N-1}{2}}\\
0 & 0 & \dotso & \sqrt{\frac{N-1}{2}} & 0
\end{array}
\right), 
\end{equation}
\begin{equation}
\label{momN}
P_{N} = -i\left(
\begin{array}
[c]{ccccc}%
0 & \frac{1}{\sqrt{2}} & 0 & \dotso & 0\\
-\frac{1}{\sqrt{2}} & 0 & 1 & \dotso & 0\\
0 & -1 & \ddots & \ddots & \vdots\\
\vdots & \dotso & \ddots & 0 & \sqrt{\frac{N-1}{2}}\\
0 & 0 & \dotso & -\sqrt{\frac{N-1}{2}} & 0
\end{array}
\right)  .
\end{equation}

Their characteristic equations are the same. Indeed, $p_{N}(\lambda) =
\det\left(  Q_{N} - \lambda I_{N}\right)  $ and $\det\left(  P_{N} - \lambda
I_{N}\right)  $ both obey the same recurrence equation:
\begin{equation}
\label{recdet}p_{N+1}(\lambda) = - \lambda p_{N}(\lambda) - \frac{N}{2}%
p_{N-1}(\lambda),
\end{equation}
with $p_{0}(\lambda) = 1$ and $p_{1}(\lambda) = -\lambda$. We have just to put
$\mathrm{H}_{N}(\lambda) = (-2)^{N} p_{N}(\lambda)$ to ascertain that the $\mathrm{H}_{N}$'s are
the Hermite polynomials for obeying the recurrence relation \cite{magnus-ob}:
\begin{align}
\label{recherm}\mathrm{H}_{N+1}(\lambda) &= 2\lambda \mathrm{H}_{N}(\lambda) - 2N \mathrm{H}_{N-1}%
(\lambda), \\
\nonumber \mathrm{H}_{0}(\lambda)& = 1, \ \mathrm{H}_{1}(\lambda) = 2 \lambda.
\end{align}
Hence the spectral values of the position operator, \emph{i.e.} the allowed or
experimentally measurable quantum positions, \underline{are} just the zeros of
the Hermite polynomials. The same result holds for the spectral values of the
momentum operator.

The non-null roots of the Hermite polynomial $\mathrm{H}_{N}(\lambda)$ form the set
\begin{align}
\nonumber Z_{\mathrm{H}}(N)&=\left\{  -\lambda_{\left\lfloor \frac{N}{2}\right\rfloor
}(N),-\lambda_{\left\lfloor \frac{N}{2}\right\rfloor -1}(N),\dotsc
,-\lambda_{1}(N),\right.\\  & \left. \lambda_{1}(N), \dotsc,\lambda_{\left\lfloor \frac{N}%
{2}\right\rfloor -1}(N),\lambda_{\left\lfloor \frac{N}{2}\right\rfloor
}(N)\right\}  ,\label{zerher}%
\end{align}
symmetrical with respect to the origin, where $\left\lfloor \frac{N}%
{2}\right\rfloor =\frac{N}{2}$ if $N$ is even and $\left\lfloor \frac{N}%
{2}\right\rfloor =\frac{N-1}{2}$ if N is odd; moreover $\mathrm{H}_{N}(0)=0$ if and
only if $N$ is odd. A vast literature exists on the characterization and
properties of the zeros of the Hermite polynomials, and many problems
concerning their asymptotic behavior at large $N$ are still open. Recent
results can be found in \cite{Elbert} with previous references therein. Upper
bounds \cite{Area} have been provided for $\lambda_{m}(N)$ and $\lambda
_{M}(N)$ where $\lambda_{m}(N)=\lambda_{1}(N)$ and $\lambda_{M}(N)=\lambda
_{\left\lfloor \frac{N}{2}\right\rfloor }(N)$ are respectively the smallest
and largest positive zeros of $\mathrm{H}_{N}$. 

However, it seems that the following observation is not known. We have studied
numerically the behavior of the product
\begin{equation}
\varpi_{N}=\lambda_{m}(N)\lambda_{M}(N)\label{lmlM}%
\end{equation}

\begin{figure}
\begin{center}
\includegraphics[width=14cm]{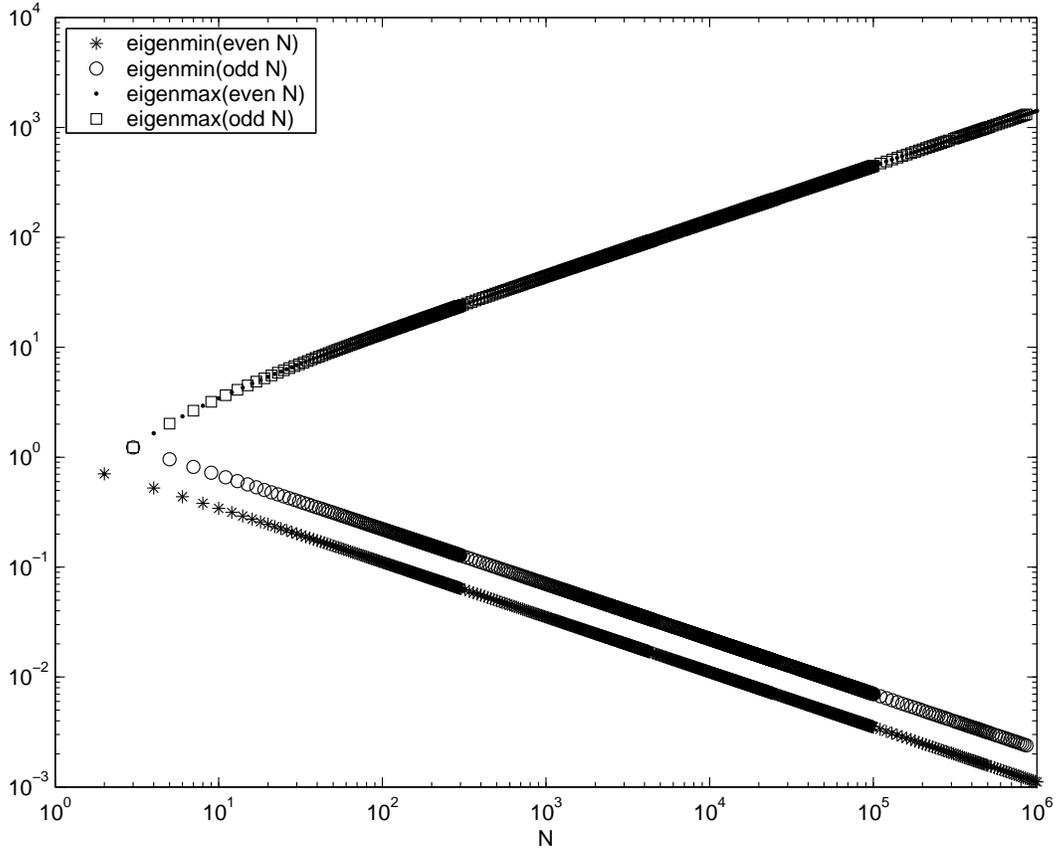}
\caption{Bottom : behaviors of the lowest positive zero eigenmin($N$) $\equiv$   $\lambda_{m}(N)$ of the Hermite polynomial of degree $N$   for
$N$ even and odd separately. Top : behavior of the largest positive zero eigenmax($N$) $\equiv$ $\lambda_{M}(N)$.}
\label{behla1}
\end{center}
\end{figure}

The zeros of the Hermite polynomials have been computed by diagonalizing the
matrix of the position operator $Q_{N}$; since $Q_{N}$ is tridiagonal
symmetric with positive real coefficients, we implemented its diagonalization
by using the QR algorithm \cite{press}; such a method enabled us to
compute the spectrum of the position operator up to the dimension $N=10^{6}$.
The respective behaviors of $\lambda_{m}(N)$ and $\lambda_{M}(N)$ are shown in Fig. \ref{behla1} for
$N$ even and odd separately. 


\begin{figure}
\begin{center}
\includegraphics[width=14cm]{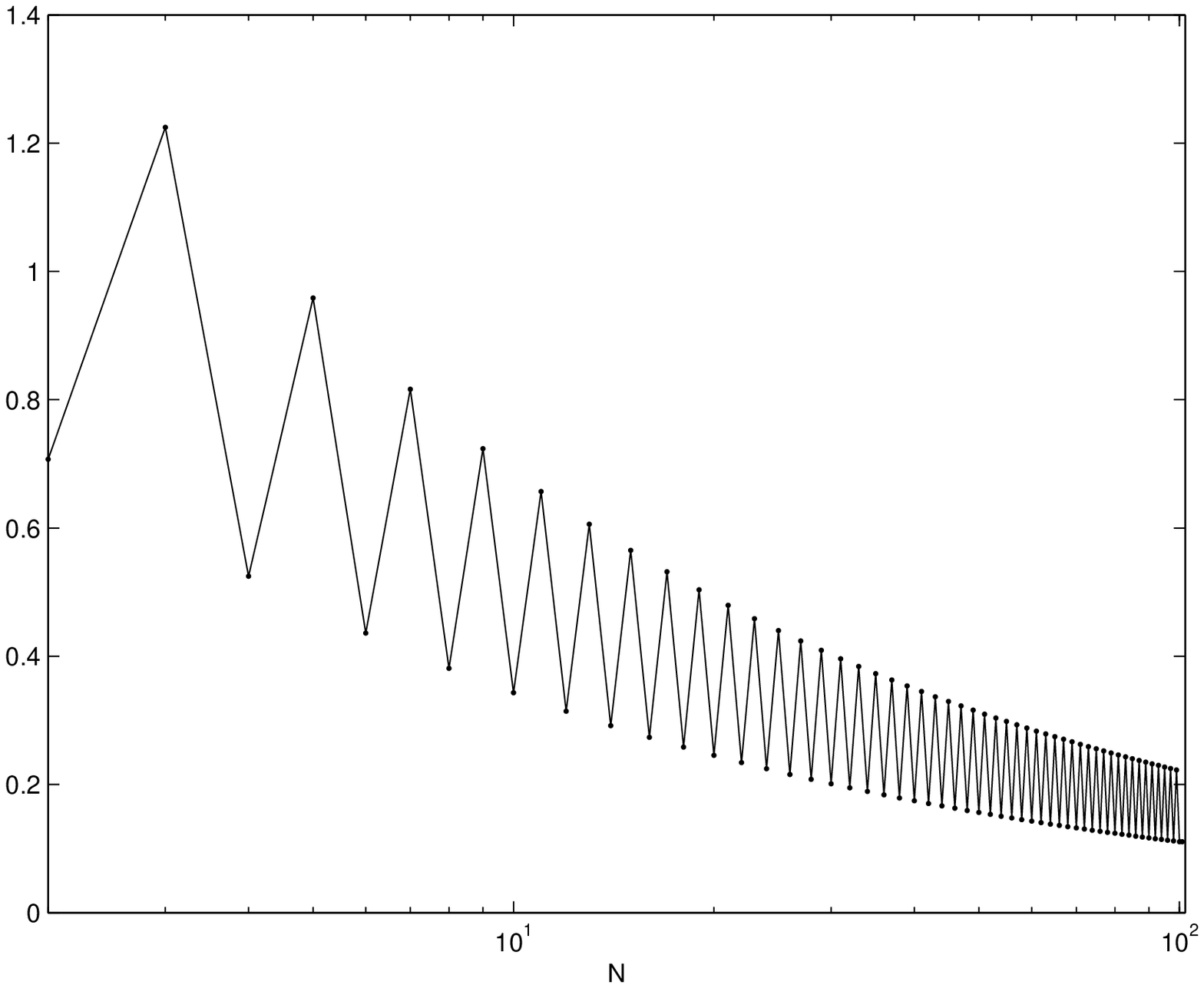}
\caption{Intertwining of the  lowest positive zeros $\lambda_{m}(N)$
of the Hermite polynomials $\mathrm{H}_{N}$ and $\mathrm{H}_{N+1}$ for small values of $N$.}
\label{zeromin}
\end{center}
\end{figure}

Now, one can easily check that $\lambda_{i+1}(N)-\lambda_{i}(N)>\lambda_{1}(N)$
for all $i\geq1$ if $N$ is odd, whereas $\lambda_{i+1}(N)-\lambda
_{i}(N)>2\lambda_{1}(N)$ for all $i\geq1$ if $N$ is even, and that the zeros
of the Hermite polynomials $\mathrm{H}_{N}$ and $\mathrm{H}_{N+1}$ intertwine, as is shown in
Fig. \ref{zeromin} in the case of $\lambda_{m}(N)$ for small values of $N$.


On the other hand, the asymptotic behavior of $\lambda_{m}(N)$ and $\lambda_{M}(N)$  can be derived from the density distribution of the  zeros of the Hermite polynomials for large $N$ \cite{lubin}. This distribution obeys the Wigner semi-circle law \cite{wigner,simisi} that gives the asymptotic behavior of the number $n(x_1,x_2)$ of zeros lying in the interval $\lbrack x_1,x_2 \rbrack$
\begin{equation}
\label{semiwig1}
n(x_1,x_2) \underset{\mathrm{large}\, N}{\approx} N \int_{x_1}^{x_2} w(t)\,  dt,
\end{equation}
with
\begin{equation}
\label{semiwig2}
 w(t) =  \frac{1}{\pi N} \sqrt{2N - t^2}.
\end{equation}
There follows from Eqs.(\ref{semiwig1},\ref{semiwig2}) that the largest zero  behaves like $\lambda_{M}(N) \underset{\mathrm{large}\, N}{\approx} \sqrt{2N}$, and the smallest positive zero behaves like $\lambda_{m}(N) \underset{\mathrm{large}\, N}{\approx} \pi/2\sqrt{2N}$ for even $N$, and  like $\lambda_{m}(N) \underset{\mathrm{large}\, N}{\approx} \pi/\sqrt{2N}$ for odd $N$.
 Hence
the asymptotical behaviors of the product $\varpi_{N} \stackrel{\mathrm{def}}{=} \lambda_{m}(N) \lambda_{M}(N)$ read respectively:%

\begin{equation}
\label{peven}
\varpi_{N}= \lambda_{m}(N) \lambda_{M}(N) \underset{\mathrm{large}\, N}{\approx} \frac{\pi}{2} ,
\end{equation}
for $ N$ even, and

\begin{equation}
\label{podd}
\varpi_{N}= \lambda_{m}(N) \lambda_{M}(N) \underset{\mathrm{large}\, N}{\approx} \pi 
\end{equation}
for $N$ odd.

Note that  (\ref{peven}) and (\ref{podd}) could as well be derived from the asymptotic values of zeros of Laguerre polynomials \cite{abram}.

Moreover, numerical studies show that the behavior of $\varpi_{N}$ is monotonically increasing for all even $N$ (resp. for all odd $N$).
Therefore, if, at a given $N$, we define by $\Delta_{N}(Q)=2\lambda_{M}(N)$
the ``size'' of the 
``universe'' accessible to exploration by the quantum system (or by the observer),
and by $\delta_{N}(Q)=\lambda_{m}(N)$ (resp. $\delta_{N}(Q)=2\lambda_{m}(N)$)
for odd (resp. even) $N$, the ``size'' of the
smallest ``cell'' forbidden to exploration by
the same system (or by the observer), we find the following upper bound for the product of these two quantities:

\begin{align}
\nonumber \delta_{N}(Q)\Delta_{N}(Q)\equiv\sigma_{N}&=\left\{
\begin{array}
[c]{ll}
4\lambda_{m}(N)\lambda_{M}(N) & \mbox{for}\ N\ \mbox{even},\\
2\lambda_{m}(N)\lambda_{M}(N) & \mbox{for}\ N\ \mbox{odd},
\end{array}
\right. \\
& \leq 2 \pi .\label{corrpos}
\end{align}

\begin{figure}
\begin{center}
\includegraphics[width=15cm]{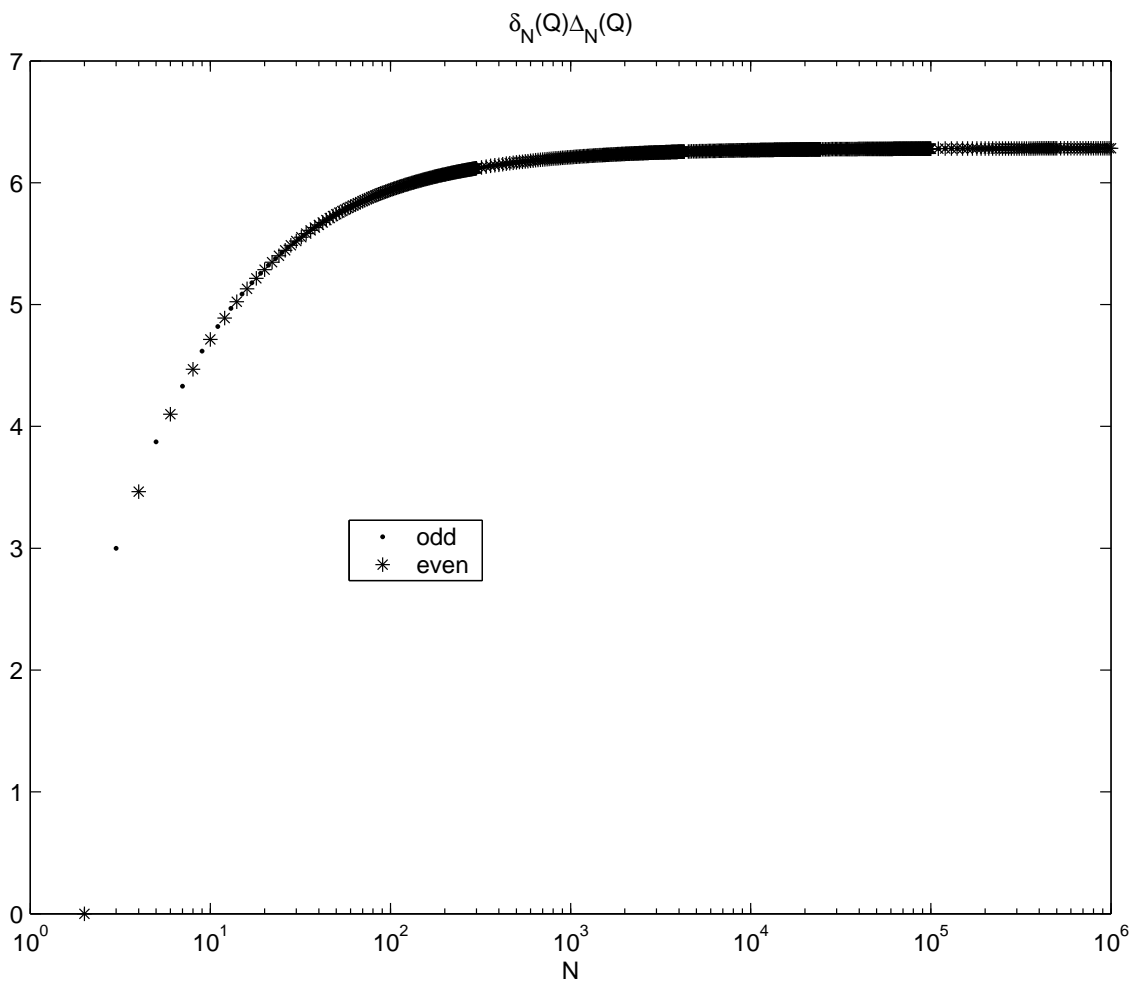}
\caption{Behavior of the product $\sigma_{N} = \delta_{N}(Q)\Delta_{N}(Q)$, as a function of $N$.}
\label{prod}
\end{center}
\end{figure}

 The monotonically increasing  behavior of the product $\sigma_{N}$, as a function of $N$,   is shown in Fig. \ref{prod} and is also given in Table \ref{tab:table1} where some values of $\sigma_{N}$ up to $N=10^{6}$ are given. . 


\TABLE[p]{
   \caption{\label{tab:table1}Values of $\sigma_{N} = \delta_{N}(Q)\Delta_{N}(Q)$ up to $N=10^{6}$. Compare with the value of $2 \pi$.}

\begin{tabular}[c]{|c|c|c|}\hline
Dimension $N$ & $\delta_{N}(Q)\Delta_{N}(Q)$& $2 \pi$\\\hline
$10$ & $4.713054$&\\ \hline
$55$ & $5.774856$&\\ \hline
$100$ & $5.941534$&\\ \hline
$551$ & $6.173778$&\\ \hline
$1$ $000$ & $6.209670$&\\ \hline
$5$ $555$ & $6.259760$&\\ \hline
$10$ $000$ & $6.267356$&\\ \hline
$55$ $255$ & $6.278122$&\\ \hline
$100$ $000$ & $6.279776$&\\ \hline
$500$ $555$ & $6.282020$&\\ \hline
$1$ $000$ $000$ & $6.282450$& 6.2831853\\\hline
\end{tabular}
}

Hence, we can assert the interesting inequality for the product (\ref{corrpos}):
\begin{equation}
\delta_{N}(Q)\Delta_{N}(Q)\leq 2\pi\ \forall N.\label{ineqpos}%
\end{equation}
Identical result holds for the momentum, of course :
\begin{equation}
\delta_{N}(P)\Delta_{N}(P)\leq 2\pi\ \forall N.\label{ineqmom}%
\end{equation}


\section{\label{sec:level5} Discussion}
In order to fully perceive the physical meaning of such inequalities, it is
necessary to reintegrate into them physical constants or scales proper to the
considered physical system, \emph{i.e.} characteristic length $l_c$  and momentum $p_c$ as was done at the beginning of Section \ref{sec:level2}:%

\begin{align}
\label{ineqposdim}\delta_{N}(Q) \Delta_{N}(Q) &\leq 2 \pi l^{2}_{c}, \\ 
\label{ineqmomdim} \delta_{N}(P) \Delta_{N}(P) & \leq 2 \pi p^{2}_{c}  \ \forall N,
\end{align}
where $\delta_{N}(Q)$ and $\Delta_{N}(Q)$ are now expressed in unit $l_{c}$.
Realistically, in any physical situation, $N$ cannot be infinite: there is an
obvious limitation on frequencies or energies accessible to
observation/experimentation. So it is natural to work with a finite although
large value of $N$, which need not be determinate. In consequence, there
exists irreducible limitations, namely $\delta_{N}(Q)$ and $\Delta_{N}(Q)$ in
the exploration of small and large distances, and both limitations have the
correlation (\ref{ineqposdim}).

Let us now suppose there exists, for theoretical reasons, a fundamental or
``universal'' minimal length, say $l_{m}$, something like the Planck length,
or equivalently a universal ratio $\rho_{u} = l_{c}/l_{m} \geq1$. Then, from
$\delta_{N}(Q) \geq l_{m}$ and (\ref{ineqpos}) we infer that there exists a
universal maximal length $l_{M}$ given by
\begin{equation}
\label{unlM}l_{M} \approx\sigma\rho_{u} l_{c}.
\end{equation}
Of course, if we choose $l_{m} = l_{c}$, then the size of the ``universe'' is
$l_{M} \approx2 \pi l_{m}$. Now, if we choose a characteristic length proper
to Atomic Physics, like the Bohr radius, $l_{c} \approx10^{-10}\mathrm{m}$,
and for the minimal length the Planck length, $l_{m} \approx10^{-35}%
\mathrm{m}$, we find for the maximal size the astronomical quantity $l_{M}
\approx10^{16}\mathrm{m}$. On the other hand, if we consider the (controversial) estimate size of our present universe $L_u = c T_u$, with $T_u \approx 13\, 10^{9}$ years \cite{age}, we get from $l_p \, L_u \approx 2 \pi l_c^2$ a characteristic length $l_c \approx
10^{-5}$m, \emph{i.e.} a wavelength in the infrared electromagnetic  spectrum...

Let us turn to another example, which might be viewed as more concrete, namely the quantum Hall effect in its matrix model version \cite{poly}. The planar coordinates $X_1$ and $X_2$ of quantum particles in the lowest Landau level of a constant magnetic field
  do not commute : 
  \begin{equation}
\label{comHall}
\lbrack X_1,X_2 \rbrack = i \theta.
\end{equation}
where $ \theta$ represents a minimal area. We recall that the average density of $N \to \infty$ electrons is related to $\theta$ by $\rho_o = 1/2\pi\theta$ and the filling fraction is $\nu = 2\pi \rho_0 / B$.
The quantity $l_m =\sqrt{\theta}$ can be considered as a minimal length. 
The Polychronakos  model deals with  finite number $N$ of electrons : 
    \begin{equation}
\label{NcomHall}
\lbrack X_{1,N},X_{2,N} \rbrack = i \theta(1 - N |N-1\rangle \langle N-1|).
\end{equation}
In this context, our inequalities read as
  \begin{equation}
\label{Hallineq}
\delta_{N}(X_{i}) \Delta_{N}(X_{i}) \leq 2 \pi l^{2}_{c}, \ i= 1,2,
\end{equation}
where $l_c$ corresponds to a choice of experimental unit.
Since $l_m =\sqrt{\theta}$ affords an  irreducible lower limit in this problem, we can assert that the 
\emph{maximal linear size} $L_M$ of the sample should satisfy :
\begin{equation}
\label{ }
l_M \leq 2 \pi \frac{l_c}{\sqrt{\theta}} l_c,
\end{equation}
for \emph{any} finite $N$.

The experimental interpretation of such a result certainly deserves a deeper investigation.

As a final comment concerning the inequalities (\ref{ineqposdim}) and (\ref{ineqmomdim}), we would like to insist on the fact they are not just an outcome of finite approximations $Q_N$ and $P_N$ (or $X_{1,N}$ and $X_{2,N}$) to the canonical position and momentum operators (or to $X_1$ and $X_2$)  in infinite-dimensional Hilbert space of quantum states. They hold however large the dimension $N$ is, as long as it is finite. 
Furthermore, let us advocate the idea that a quantization of the classical phase space results from the  choice of a specific (reproducing) Hilbert subspace $\mathcal{H}$ in $L^2(\mathbb{R}^2, d\mu(q, p))$ in which coherent states provide a frame resolving the identity. This frame corresponds to a certain point of view in dealing with the classical phase space, and this point of view yields the quantum versions $Q_N$ and $P_N$ (or $X_{1,N}$ and $X_{2,N}$) of the classical coordinates $q$ and $p$ (or $x_1$ and $x_2$).


\begin{acknowledgments}
 The authors are indebted to   Alphonse Magnus, Jihad Mourad and Andr\'e Ronveaux for  valuable comments and discussions.
\end{acknowledgments}


\end{document}